\documentclass[useAMS,usegraphicx,usenatbib]{mn2e}
\usepackage{times}

\title[Cement dust]{Do cement nanoparticles exist in space?}
\author[G. Bilalbegovi\' c et al]
{G. Bilalbegovi\' c$^{1}$,
A. Maksimovi\' c$^{2}$,
V. Moha\v cek-Gro\v sev$^{2}$
\\
$^{1}$Department of Physics, Faculty of Science, University of Zagreb, Bijeni\v cka 32, 10000 Zagreb, Croatia\\
$^{2}$Rudjer Bo\v skovi\' c Institute,
Bijeni\v  cka 54, 10000 Zagreb, Croatia
}
\begin{document}

\date{\today}

\pagerange{\pageref{firstpage}--\pageref{lastpage}}
\pubyear{2013} \volume{000}

\maketitle 
\label{firstpage}

\begin{abstract}
The calcium-silicate-hydrate is used to model  properties of cement on Earth.
We study cementitious nanoparticles and propose these structures as components of cosmic dust grains. 
Quantum  density functional theory methods are applied  for the calculation of infrared spectra of  Ca$_4$Si$_4$O$_{14}$H$_4$, 
Ca$_6$Si$_3$O$_{13}$H$_2$,  and Ca$_{12}$Si$_6$O$_{26}$H$_4$  clusters. 
We find bands distributed over the near, mid and far-infrared region.
A specific calcium-silicate-hydrate spectral feature at 14 $\mu$m, together with the bands 
at 10  and 18  $\mu$m which exist for other silicates as well, 
could be used for a detection of cosmic cement.
We compare calculated bands with the 14 $\mu$m features in the spectra of HD 45677, HD 44179, and IRC+10420 which were observed by  {\it Infrared Space
Observatory} and classified as remaining.
High abundance of oxygen atoms in cementitious nanoparticles could partially explain observed depletion of this element from the interstellar medium
into dust grains.
\end{abstract}


\section{Introduction}
\label{intro}

Dust  is  produced in supernova explosions and the outflows of  stars \citep{Draine2011,Tielens2013}.  It is  very important in the evolution of stars and galaxies.
Dust grains act as a substrate for the formation of H$_2$ and many other astrophysical molecules, including prebiotic ones.
It is accepted today that cosmic dust consists of silicates and carbonaceous materials with impurities from several  chemical elements.
Using infrared spectroscopy both silicates and carbon based grains have been observed 
in various environments:  close to Earth and exoplanets, 
in the interstellar medium,  in comets,  around stars, and active galactic nuclei.
The light from cosmic objects heats dust grains, which is followed by a dust emission in the infrared spectral region.
Absorption features in the infrared produced by dust are also observed.
The size of grains is from  the length of one molecule, up to several hundred micrometers.  
It is known that 10 per cent of grains are ultra-small, with the  size around 1.5 nm or less \citep{Li2001}. 

Cosmic dust mainly forms from the most abundant elements, such as silicon, carbon, oxygen, iron, and magnesium.
Calcium is one of chemical elements that are produced in supernova explosions.  It belongs to the group of twenty most abundant chemical elements in the Universe. 
The Ca II  absorption was one of the first lines discovered in the interstellar space \citep{Hartmann1904}.  
It has been proposed that compounds 
which contain calcium  exist in space. For example, spectra of
Ca-poor and Ca-rich pyroxenes have been measured in the mid and far-infrared wavelengths  in laboratories on Earth \citep{Koike2000}.
It has been suggested that  the amorphous diopside,  CaMgSi$_2$O$_6$, contributes to the far-infrared spectrum of the planetary nebulae NGC 6302.
Calcium has been measured in meteorites and rocks of Mars where it is a component 
in the calcium-aluminum-rich inclusions and chondrules \citep{Simon2009}.

Calcium compounds are the main constituents of cement, which 
is one of the most used materials on Earth \citep{Allen2007,Pellenq2009,Skinner2010,Masoero2012}.
The cement paste is formed when the cement powder, consisting mainly of alite (Ca$_3$SiO$_5$) and belite (Ca$_2$SiO$_4$),
 is mixed with water.    The  calcium-silicate-hydrate
has a granular complex structure.  Its chemical composition is variable, being characterized by the CaO:SiO$_2$ ratio in the range from 0.7 to 2.3 \citep{Manzano2007}. It has been found that temperatures above 350 K are necessary for the reaction between H$_2$O gas molecules and silicates in the outflows of stars \citep{Grossman1974}. However,
silicate cosmic dust grains are often covered with water  ice mantles. H$_2$O from mantles could react with silicates and calcium atoms 
and form a cosmic cement paste.  
We expect that cementitious materials could be produced as a dust component around oxygen-rich stars where silicates are dominant. 
The presence in space of various minerals in the form of hydrous silicates (for example, talc Mg$_3$[Si$_4$O$_{10}\mid$(OH)$_2$] and
montmorillonite (Mg,Al)$_2$[Si$_4$O$_{10}\mid$(OH)$_2$](Na,K,Ca)$_x\cdot$nH$_2$O)  has been under discussion for some time \citep{Whittet1997,Hofmeister2006,Mutschke2008}.

Infrared spectroscopy is used  in the field of solid state astrophysics  \citep{Jager2011},
and astronomical observations are compared with spectra measured in laboratories.
For example, spectra between 2.4   and 195 $\mu$m  for 17 oxygen-rich circumstellar dust shells 
were observed using the Short and Long Wavelength Spectrographs on the {\it Infrared Space
Observatory} ({\it ISO}) and compared with  laboratory measurements \citep{Molster2002a,Molster2002b}.
Although many bands were fitted with  measured spectra of Mg-rich olivines and pyroxenes, 20 per cent of the spectral features were not identified.  The {\it Spitzer} and {\it Herschel} missions collected many infrared spectra of dust grains
\citep{Watson2009,Ciesla2014}.
In laboratory measurements, in order to model various cosmic conditions,
spectra of materials are studied  from very low temperatures up to 1000 K.
However,  the interpretation of measured infrared spectra of complex materials can
often be involved.  Infrared spectroscopy of nanoparticles is even more demanding in comparison
with measurements of the bulk material.
Computational infrared studies of dust materials provide a connection between the microscopic structure and spectral properties. 
This is very important in astrophysics. For example,
because of astronomical emissions at 3.3, 6.2, 7.7, 8.6,  11.2,  and 12.7 $\mu $m,
infrared spectra of polycyclic aromatic hydrocarbons (PAHs)  have been calculated and assembled in a spectroscopic database \citep{Bauschlicher2010,Boersma2014}.
Recently, spectra of  PAHs clusters have been calculated \citep{Ricca2013}.
Infrared spectra of a dust grain model in the form of the
 bare nanopyroxene cluster Mg$_4$Si$_4$O$_{12}$, as well as  of its hydrogenated and oxygenated  forms, 
have also been calculated by quantum computational methods
\citep{Goumans2011}. 

Various  crystalline and amorphous forms of silicates with different sizes, shapes and
chemical compositions could form
under diverse conditions in space \citep{Henning2010}.
We have selected three clusters which consist of Ca, Si,  O, and H atoms.
Using these nanoparticles  we model the crystalline and amorphous states of
possible ultra-small cosmic dust grains with the structure of cement paste. 
Density functional theory computational methods \citep{Martin2004} are used to study structural properties of these nanoparticles,  calculate vibrational modes,
and explore important features in their infrared spectra.

\section{Computational methods}
\label{methods}

The accepted model of cosmic silicates is an arrangement of small silicate particles of different sizes \citep{Henning2010}.
We study three nanoparticles which exhibit the typical bonding and small scale structure of the cement paste. 
These three models of ultra-small cosmic particles are selected to represent
crystalline and amorphous cosmic silicates, both found in cosmic dust. 
One cluster, Ca$_4$Si$_4$O$_{14}$H$_4$ (labelled as C1 in the following text), is chosen to model a crystalline state of any possible cosmic cement. 
This cluster has been modelled on the previous Hartree-Fock calculation on the formation of cementitious nanoparticles  \citep{Manzano2007}. 
The structure of  C$_1$ has been proposed by Manzano et al as a common precursor of  both  tobermorite  and jennite  bulk crystal  structures of the  calcium-silicate-hydrate. The same C$_1$ structure, but in the narrower spectral region, has been studied using the
semi-empirical MNDO and density functional theory methods \citep{Bhat2011}.
We also study two amorphous nanoparticles, Ca$_6$Si$_3$O$_{13}$H$_2$ (C2 in the following text) and Ca$_{12}$Si$_6$O$_{26}$H$_4$ (labelled as C3). They 
have been cut from the bulk of the calcium-silicate-hydrate presented in the Materials Project  \citep{Jain2013}. 
The clusters are characterized by the CaO:SiO$_2$ ratio of 1 (in C1)
and 2 (in C2 and  C3).  One (in C2) and two (in C1 and C3)  units of water (2H and O) are present.

We have used the real-space GPAW density functional program package \citep{Enkovaara2010} and its ASE (Atomistic Simulation Environment) user interface
\citep{Bahn2002}.
The PBE generalized gradient approximation (GGA) exchange-correlation functional  \citep{Perdew1996}, and
the PAW (projector augmented wave) pseudopotentials \citep{Mortensen2005}
are applied. The C1 cluster  is globally optimized.  We use the  Monte Carlo basin-hopping algorithm \citep{Wales1997,Wales1999} 
as implemented in ASE. All interactions in this optimization are treated within GPAW,  at the same density functional theory level
(i.e., using the GGA functional and the PAW pseudopotentials).
It is known that the  Monte Carlo basin-hopping algorithm
method is very successful in finding minima of clusters \citep{Bilalbegovic2003,Goumans2011,Jiang2011}.
Infrared spectra are calculated using the finite difference approximation, both for
 the dynamical matrix and the gradient of the dipole momentum of the system
\citep{Porezag1996,Frederiksen2007}. In this method the infrared intensity $I_i$ of the mode $i$ is calculated from
\begin{equation}
  I_i = \frac{{\cal N}\pi}{3c}{\left |{\frac{d{\vec{\mu}}}{dQ_i}}\right |}^2, 
\label{eq1}  
\end{equation}
where $\cal N$ is the particle density, $c$ is the velocity of light, $\mu$ is the electric dipole momentum, and $Q_i$ is the coordinate of the normal mode.
The GPAW code and the PBE GGA functional  have  been recently  employed to calculate the infrared spectrum of thiol-stabilized gold nanoclusters \citep{Hulkko2011}. These results have been compared with measurements
and  a very good agreement is found.
Our results for the crystalline C1 cluster agree with the calculations of Bhat and Debnath, which have been carried out at local density approximation (LDA) and GGA BLYP levels \citep{Bhat2011}. Two bands in
spectra of amorphous clusters C2 and C3
agree with laboratory measurements on calcium silicate smoke samples in the mid-infrared region \citep{Kimura2005}.

\section{Results and Discussion}
\label{results}

\begin{figure}
\centering 
\includegraphics[width=5.5cm]{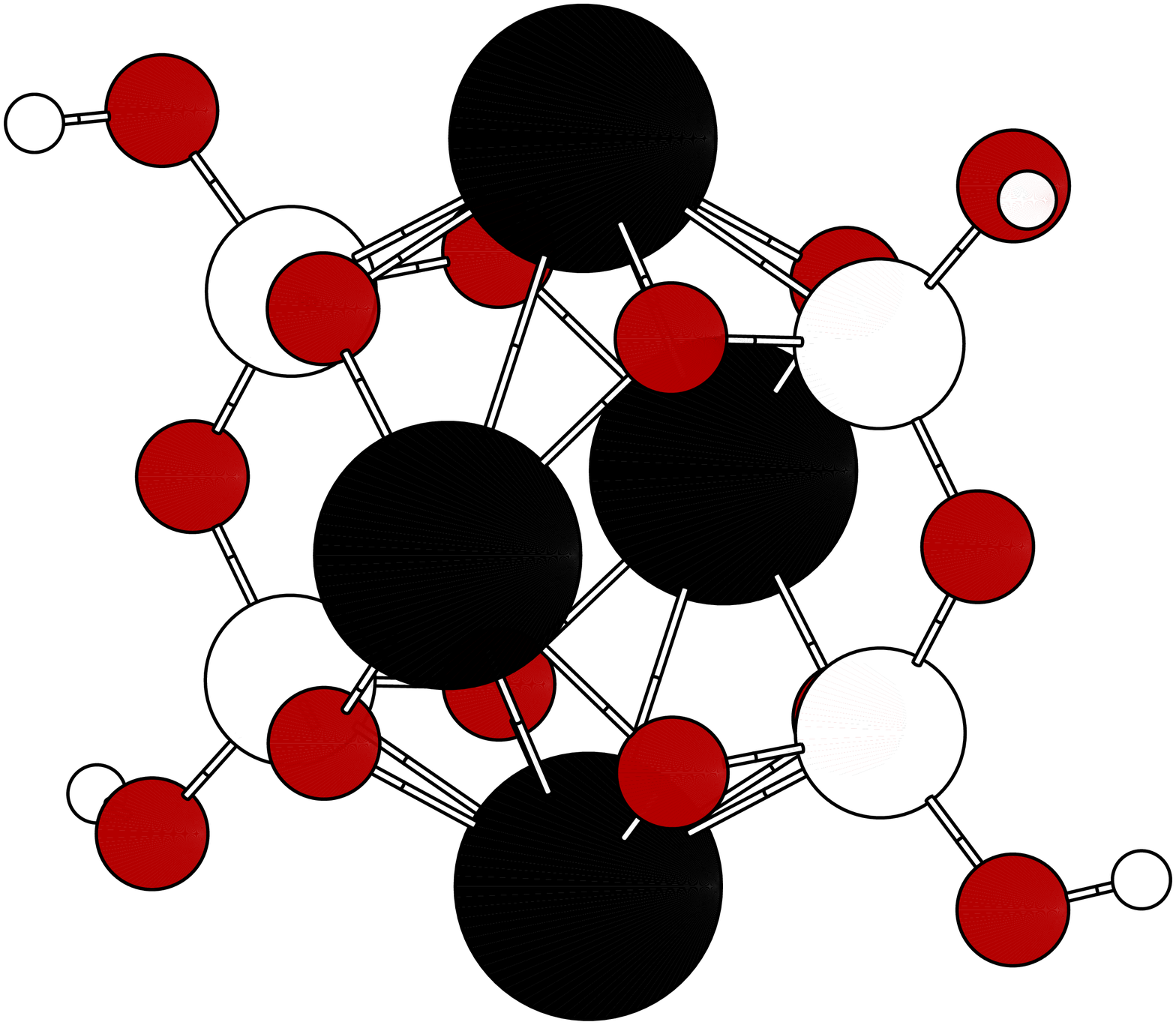}
\includegraphics[width=6.5cm]{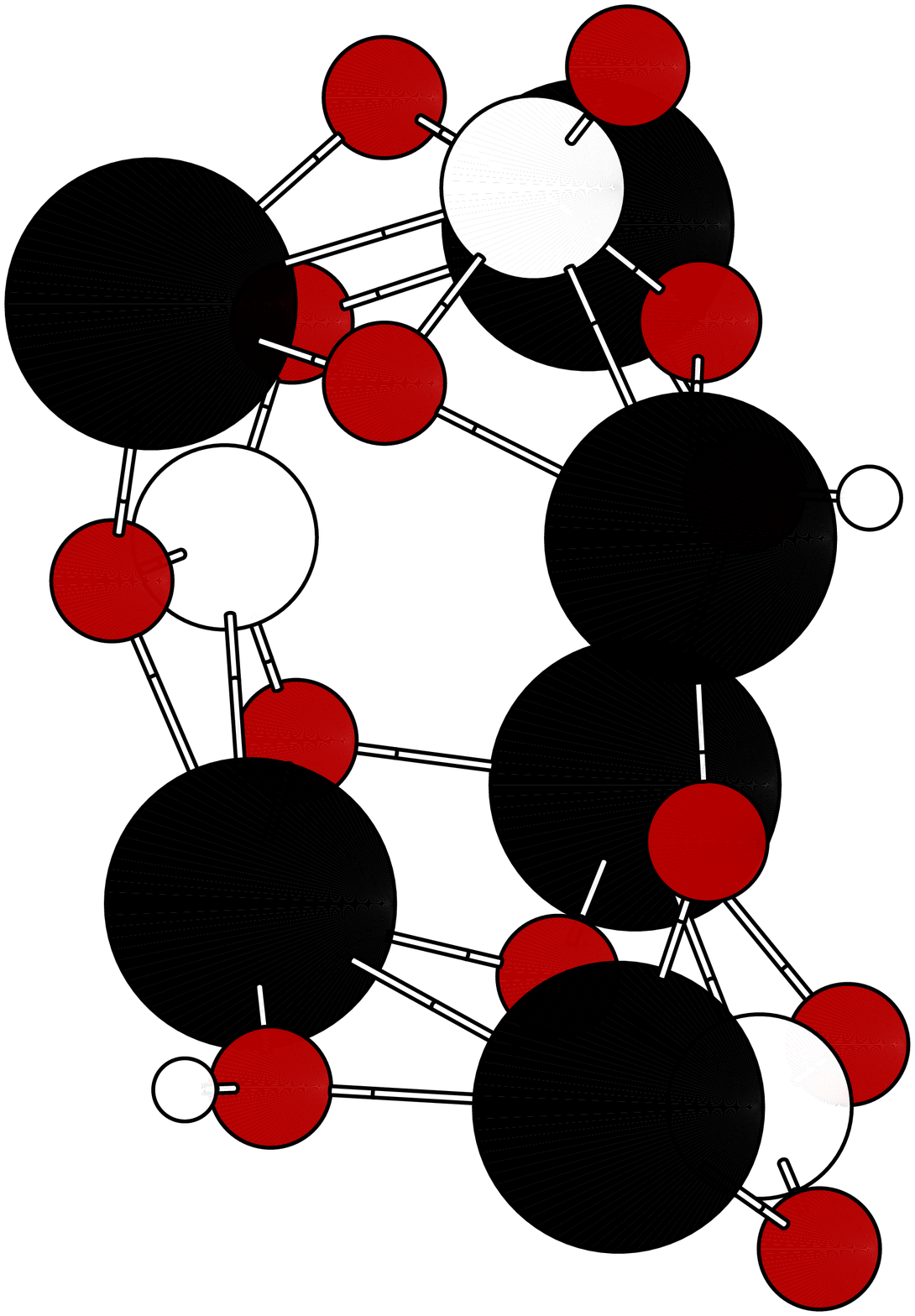}
\includegraphics[width=6.5cm]{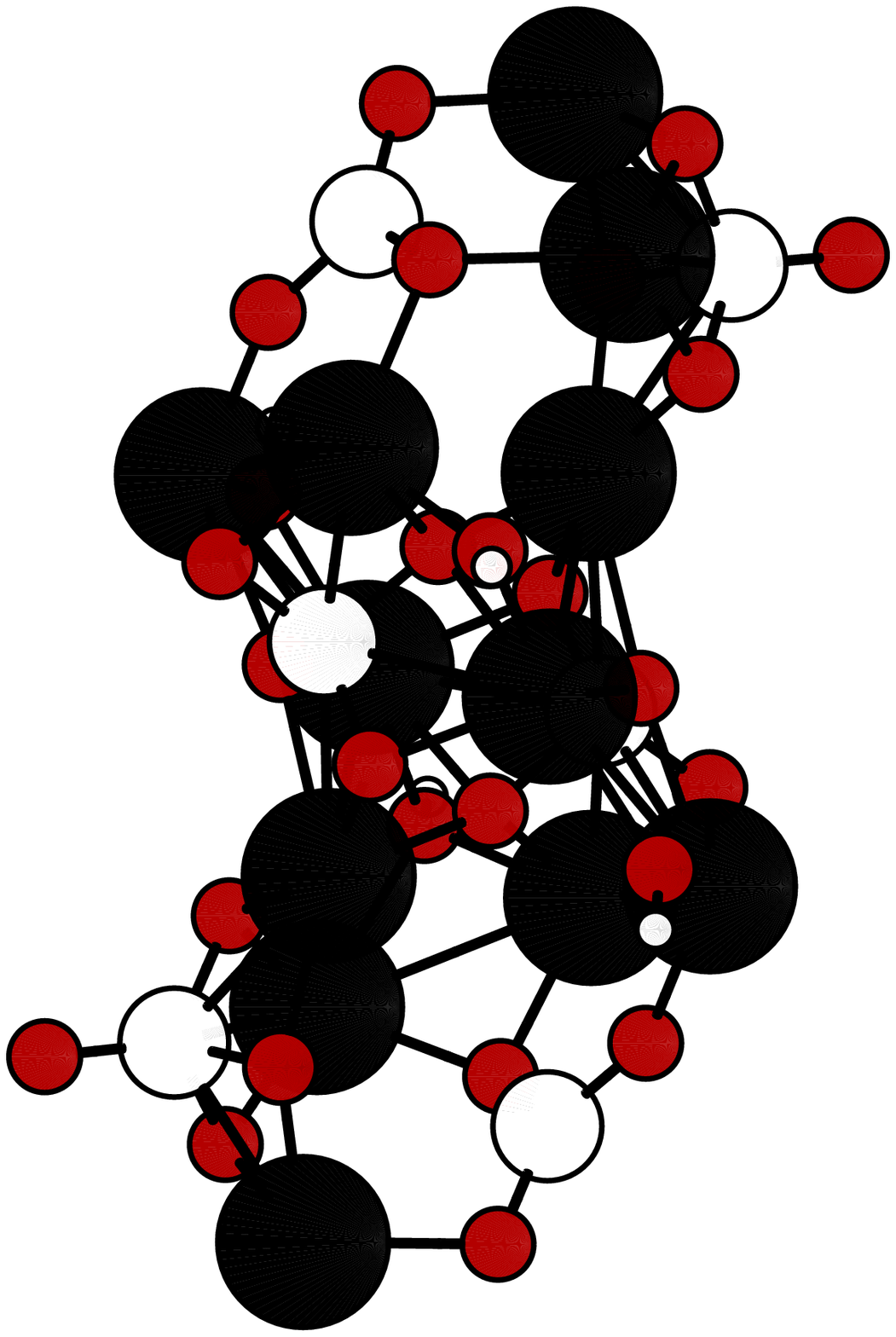}
\caption{
The geometry of cementitious clusters, (a) C1:  Ca$_4$Si$_4$O$_{14}$H$_4$, (b) C2: Ca$_6$Si$_3$O$_{13}$H$_2$, (c) C3: Ca$_{12}$Si$_6$O$_{26}$H$_4$. 
Small white balls represent hydrogen atoms,  gray (red in the  color figure)  are for oxygen, large white for silicon, and black for calcium atoms.}
\label{fig1}
\end{figure}

Optimized structures for the cementitious clusters we have studied are shown in Fig.~\ref{fig1}.
We have found that clusters C2  and C3  are less stable than the crystalline precursor
C1. A small displacement of atoms from the structures of C2  and C3, corresponding to the their local minima,
quickly leads to bigger forces between atoms.  
Potential energies are: -6.771 eV/atom (C1),  -6.668 eV/atom (C3), and -6.476 eV/atom (C2). 
When compared to C2, the cluster C3 has all numbers of atoms doubled.
However, C2 is less stable than C3. 
Clusters C2  and C3 represent amorphous  nanoparticles that could form under various conditions of temperatures, pressures, radiations, and shock waves in space.
In contrast, the structure of
C1 is very stable - which confirms the idea of  Manzano et al  \citep{Manzano2007}
that this cluster is a building block of cement on Earth. 
Clusters C2 and C3 are examples of many possible amorphous silicate nanoparticles in space.
We have calculated that, in addition to the specific cement band at 14 $\mu$m, amorphous clusters C2 and C3 exhibit the typical bands of amorphous cosmic silicates at 10 and 18 $\mu$m. This shows that our models of amorphous cement nanoparticles are reliable.
Both amorphous and crystalline silicates have been observed in cosmic dust  \citep{vanBoekel2005}. 
Amorphous silicates are typical for the interstellar medium, whereas in circumstellar disks both crystalline and amorphous silicates exist.  Crystalline silicates are also found in comets and dust particles in the Solar system. Infrared spectra are used for the analysis of these various forms of silicate cosmic dust  \citep{Kessler2006,Henning2010,Draine2011}.

In the majority of bulk silicates the Si atom is positioned in the tetrahedra of oxygen atoms.  In the mineralogy of silicates several complex structures are known, for example two tetrahedra sharing one oxygen atom (as in gehlenite Ca$_2$Al$_2$SiO$_7$),
or tetrahedra that form rings (as in the beryl Be$_3$Al$_2$Si$_6$O$_{18}$). Several other arrangements of chains and rings exist in various silicates.  Hydrous silicates belong to the group of phyllosilicates where layers of silicates and OH layers appear.
Higher coordination numbers in silicates are also known. For example, the octahedral configuration of Si occurs under high pressures, or in the mineral thaumasite,
Ca$_3$Si(OH)$_6$(CO$_3$)(SO$_4$)$\cdot$12H$_2$O. Thaumasite sometimes forms in cements when their main components are exposed to sulphates \citep{Nobst2003}.
We have found that in the C1 cluster all silicon atoms are tetrahedrally coordinated. In  C2 two Si atoms are surrounded by three oxygens, and one with four O atoms. Four Si atoms are tetrahedrally coordinated, and two silicon atoms have three nearest oxygen
neighbours in the C3 cluster. 

\begin{figure}
\centering 
\includegraphics[width=9.0cm,height=6cm]{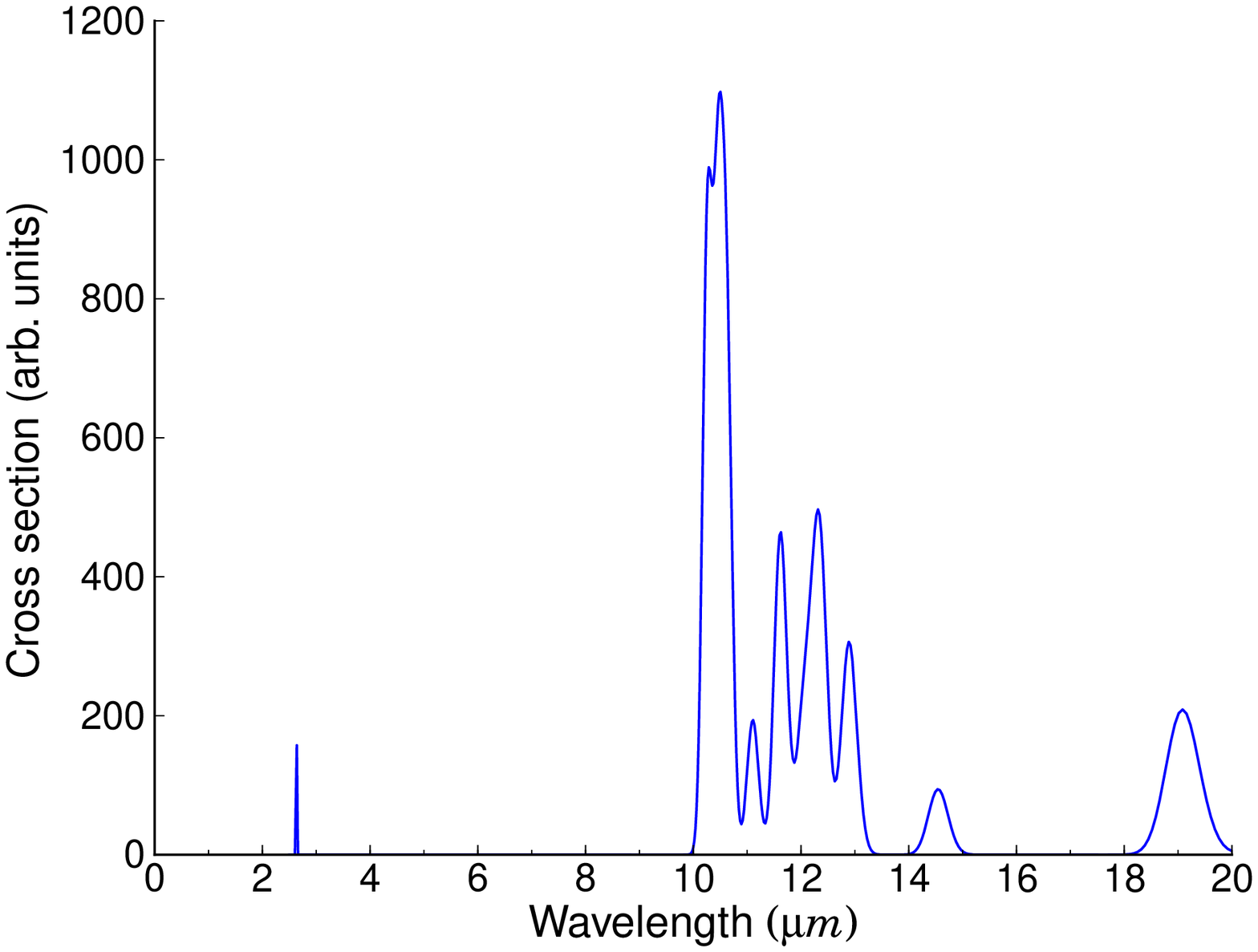}
\includegraphics[width=9.0cm,height=6cm]{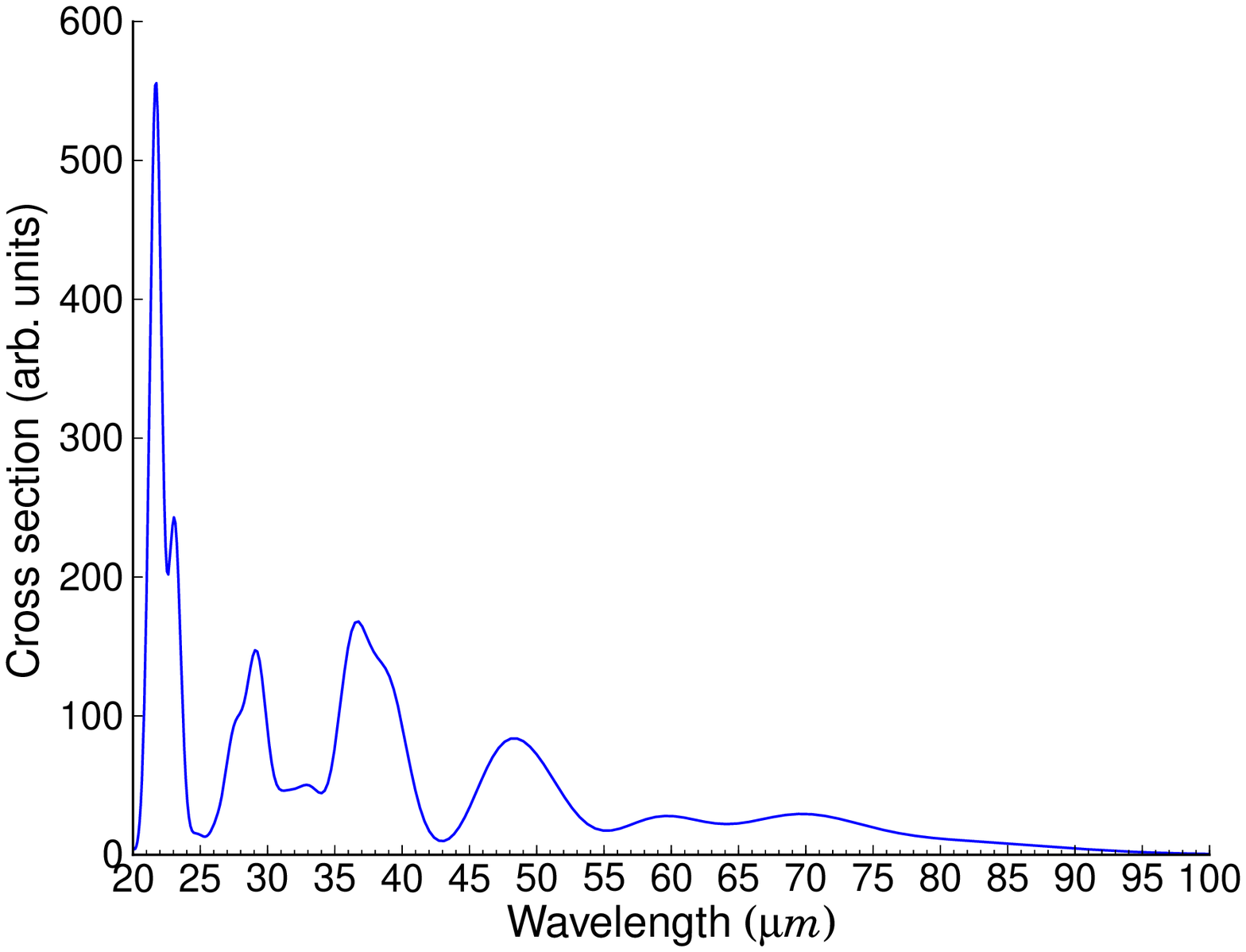}
\caption{Infrared spectrum  of the C1 cluster.}
\label{fig2}
\end{figure}

\begin{figure}
\centering 
\includegraphics[width=9.0cm,height=6cm]{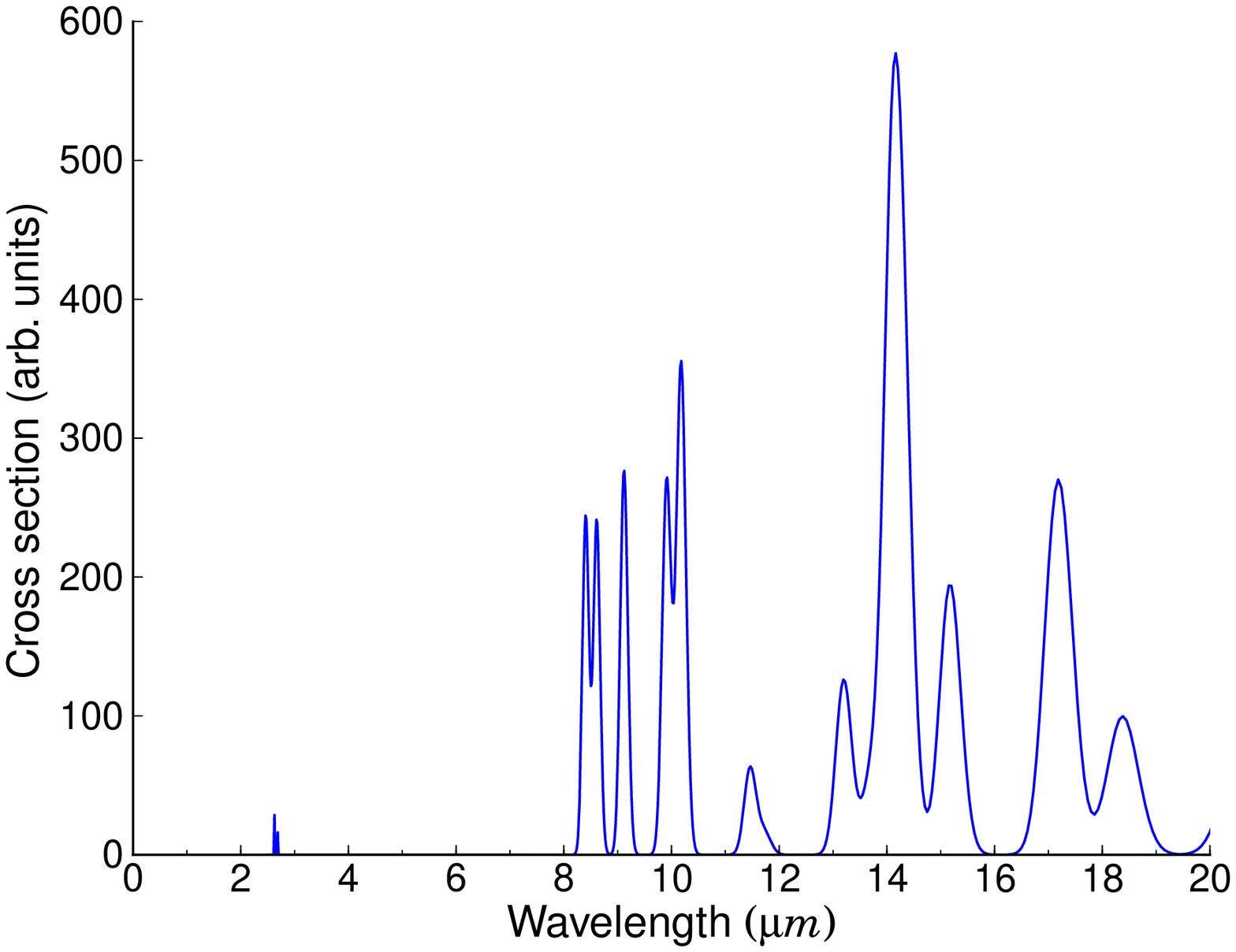}
\includegraphics[width=9.0cm,height=6cm]{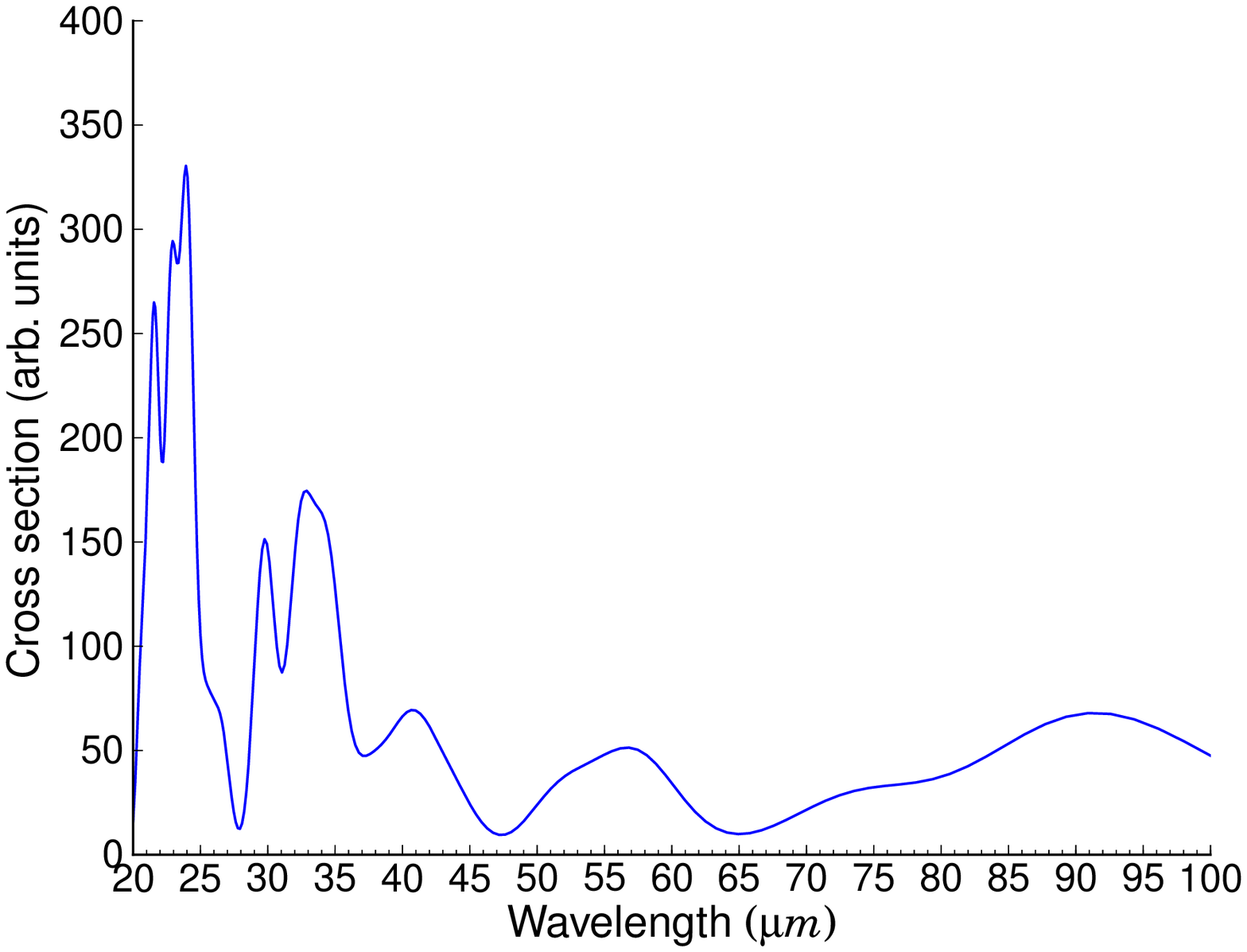}
\caption{Infrared spectrum of the  C2 cluster.}
\label{fig3}
\end{figure}

\begin{figure}
\centering 
\includegraphics[width=9.0cm,height=6cm]{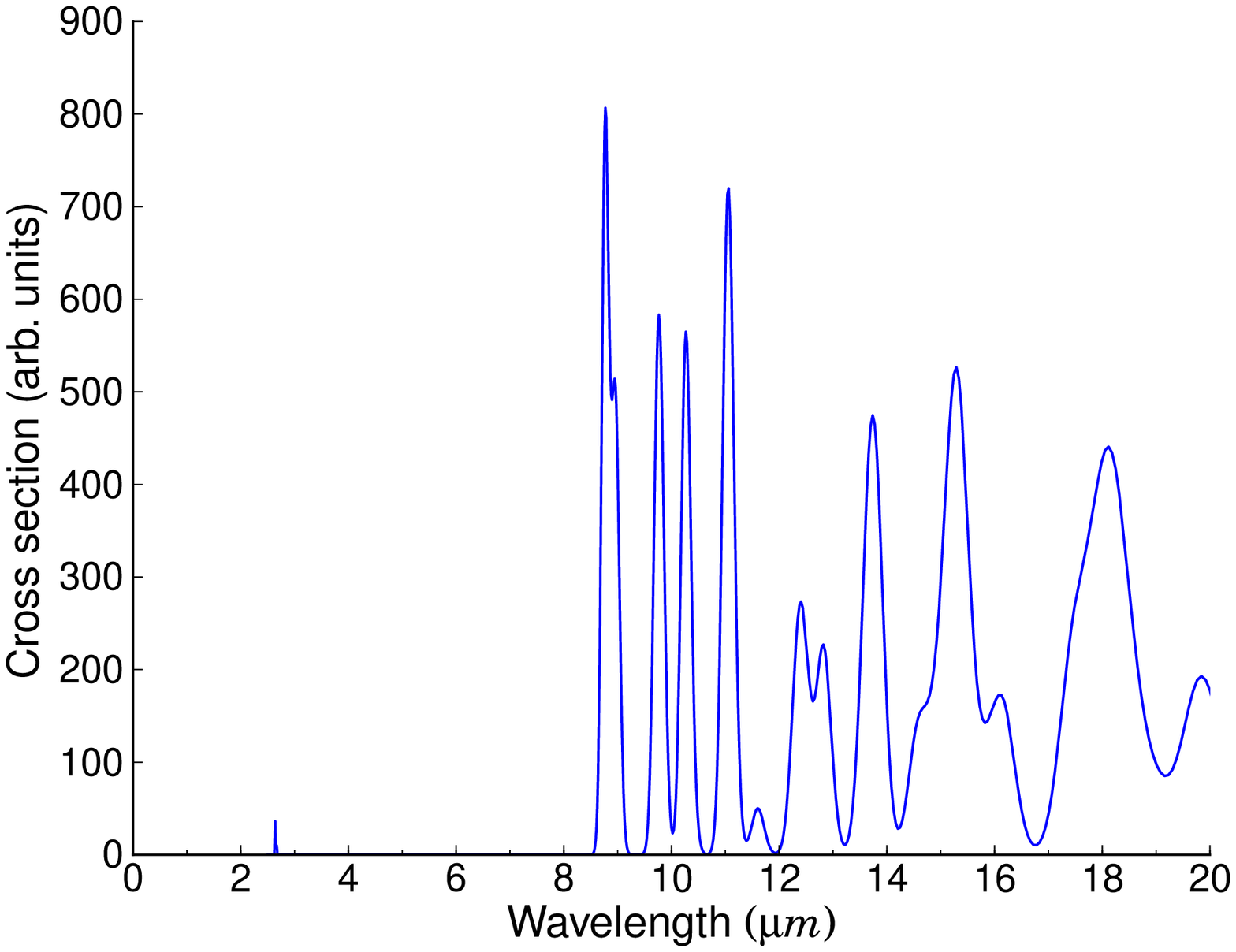}
\includegraphics[width=9.0cm,height=6cm]{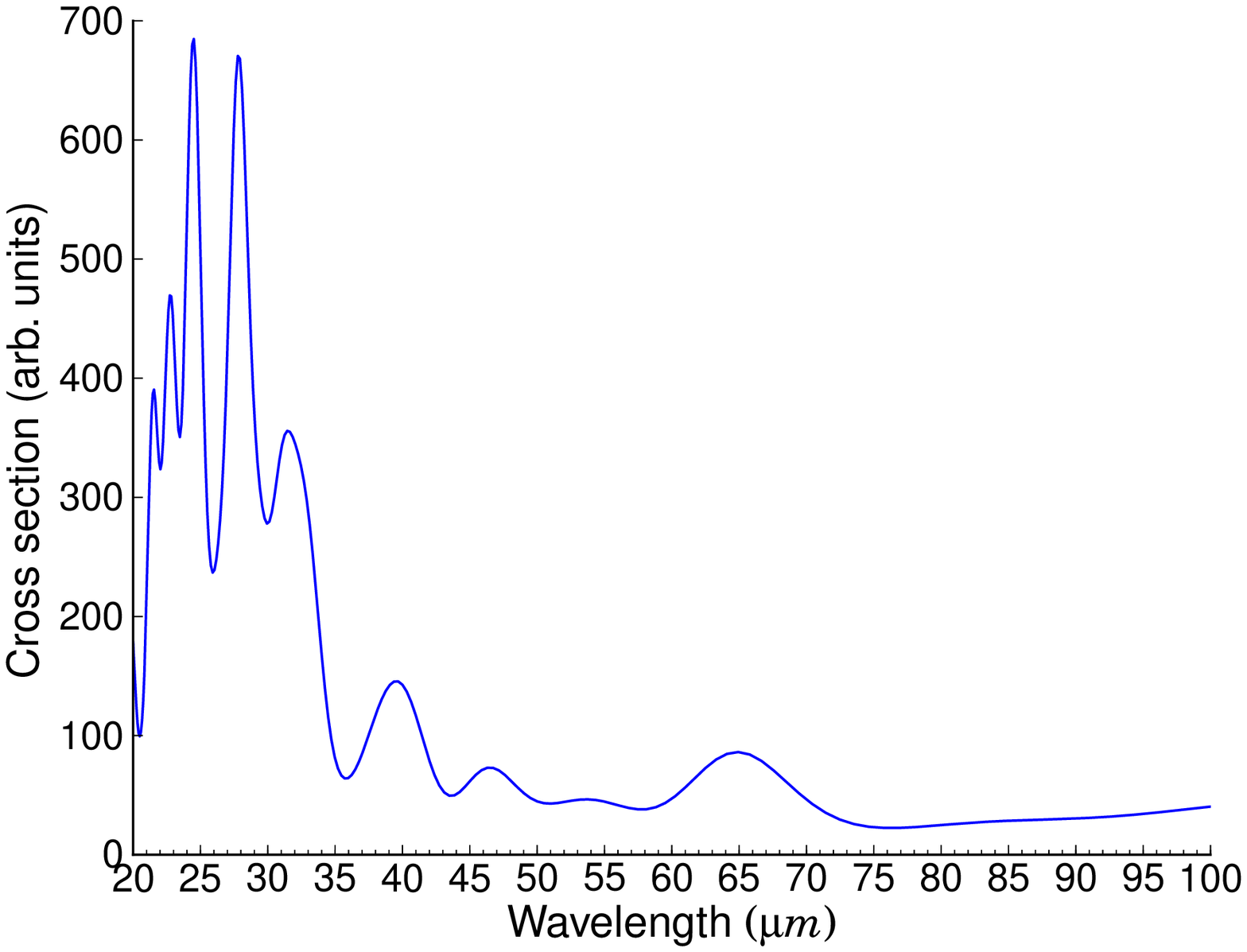}
\caption{Infrared spectrum of the  C3 cluster.}
\label{fig4}
\end{figure}

Infrared spectra (calculated  for nanostructures presented in  Fig.~\ref{fig1}) are shown in Figs.~\ref{fig2}-\ref{fig4}.
Theoretical results for infrared spectra consist only of vertical lines. In contrast, in measured astronomical IR spectra these lines are broad
and reflect conditions of observed objects \citep{Bauschlicher2010}.  It is a common practice to broaden theoretical lines.
We use  the Lorentzian band profiles with FWHM of 20 cm$^{-1}$.
Bands with higher intensities, calculated using Eq.\ref{eq1}, are shown in Table \ref{tab1}.

The lines with a small intensity exist in the near-infrared region at  2.62, 2.64, and 2.65 $\mu$m for C1 (Fig.~\ref{fig2}),
2.63   and  2.69 $\mu$m for  C2 (Fig.~\ref{fig3}), 2.64,  2.66, and 2.67 $\mu$m
for C3 (Fig.~\ref{fig4}).
Bhat and Debnath for C1 calculated 2.63 (in LDA) and 2.67 $\mu$m (in GGA BLYP) \citep{Bhat2011}.
A spectral region at $\sim$(2 -- 4) $\mu$m is known as the water ice band \citep{Gibb2004}. The interval (2.6 -- 2.8) $\mu$m corresponds to vibrations of O-H in hydrated silicates.
The water ice band has been studied in the astrophysical context because of various hydrosilicate minerals \citep{Whittet1997,Hofmeister2006,Mutschke2008}, or H$_2$O trapped and adsorbed in the SiO condensate \citep{Wada1991}.

Bands in the near and beginning of mid-infrared region have been  found in laboratory measurements of CaO and Ca(OH)$_2$ grains and 
the measured line at 6.8 $\mu$m has been proposed to
explain this feature in the spectra of young stellar objects \citep{Kimura2005}.
Typical features of amorphous silicates in mid-infrared are broad bands at 10 and 18 $\mu$m which correspond to the Si-O stretching and O-Si-O bending modes. Silicate bands in the far infrared also exist and they are attributed to metal-oxygen modes. 

In the mid-infrared region of the C1 cluster (Fig.~\ref{fig2} and Table~\ref{tab1}) we found five close bands at 10 $\mu$m: 
10.26, 10.29, 10.40, 10.47,  and 10.6{\bf 3} $\mu$m. In addition, we have calculated several bands of a smaller intensity
between 11  and 13 $\mu$m,  as well as  at 14.44, 14.55  and 19.08 $\mu$m. Several bands appear above 20 $\mu$m, the strongest ones at 21.73
and 23.11 $\mu$m. The far-infrared spectrum is broad and with a small intensity. Main bands in this region are 36.44  and 38.76 $\mu$m.
Bhat and Debnath for the mid-infrared region of C1 also
did not find any lines below 10 $\mu$m in their GGA BLYP approximation, but such lines appear in LDA \citep{Bhat2011}. Our other calculated lines in mid-IR are in good agreement with their results. They did not calculate bands above 22 $\mu$m.

In the mid-infrared spectral region of the  C2 cluster (Fig.~\ref{fig3} and Table~\ref{tab1}) bands appear also slightly below 10 $\mu$m: 8.41, 8.61, 9.12, 
and 9.91 $\mu$m. There are strong bands at 10.18, 14.32  and 17.18 $\mu$m. 
Several other bands of a smaller intensity exist between 10  and 20 $\mu$m. For the C2 cluster
many bands appear above 20 $\mu$m. The strongest bands in this region are: 22.90, 24.03, 29.9, and 34.47 $\mu$m.
In the mid-infrared spectral region of the  C3 cluster (Fig.~\ref{fig4} and Table~\ref{tab1})
stronger bands are  close to 8, 9, 10, 11, 12, 13, 14, 15, 16, 17, 18, 19, 21, 22, 24, 27, 29, and 30  $\mu$m. Weaker bands exist above 30  $\mu$m.
A rather strong band (with the intensity of  71.98 km/mol) is calculated at 64.93 $\mu$m.
The calcium-silicate compound diopside CaMgSi$_2$O$_6$, proposed as a contributor to the far-infrared spectrum of the planetary nebulae NGC6302,
shows the band at 65 $\mu$m  \citep{Koike2000}.

\setcounter{table}{0}
\begin{table*}
\centering
\caption{Infrared bands for the C1, C2 and C3 clusters. Only bands with intensities above 70 km/mol are shown for C1, above 60 km/mol for C2, and above 130 km/mol for C3.}
\label{tab1}
  \begin{tabular}{ c| c | c|c|c|c|}
    \hline
     C1      &  & C2 &  & C3 \\
    \hline
    Band  ($\mu$m) & Intensity (km/mol) &  Band  ($\mu$m) & Intensity (km/mol)  &  Band  ($\mu$m) & Intensity (km/mol) \\
  \hline 
  2.64    &  70.19 &  8.41 &  243.78 &   8.77   &  493.45  \\ 
  10.26   &  820.40 & 8.61 &  240.98 &    8.92   &  264.52   \\  
  10.40   &   93.81 &  9.12 &  277.37 &   8.99   &  282.56  \\   
  10.47   &  786.97 &  9.91 &  269.79 &   9.74   &  382.20  \\   
   10.63  &   671.12 & 10.18 & 354.15 &   9.81   &  262.97 \\  
   11.11  &   183.01 & 11.46 &  62.57 &   10.27   & 558.72   \\ 
   11.61  &  136.73 & 13.20  &  126.10 &  11.05   &  721.71  \\  
   11.62  &  249.77 & 14.10 &  434.48 &   12.34  &  153.84  \\ 
   12.04  & 178.55 &   14.32 &  257.23 &  12.46  &  152.52  \\  
   12.23 &  179.10 &   15.17 &  195.06 &  12.82 &   221.98\\     
   12.35  & 127.78 &  17.18  & 252.09 &   13.63  &  183.52  \\ 
   12.38  &  240.04 &  18.38  &  99.73 &  13.79  &  343.60 \\    
   12.89  &  291.28 &  20.60  &  64.24 &  14.58  &  130.18 \\    
   14.55  &  89.18 &   21.53  &  106.24 &  15.23  &  282.64 \\   
   19.08  & 208.14 &  21.64  &  140.84 &  15.38  &  252.60 \\  
   21.22  & 99.59 &   22.90  &  225.48 &  16.16  &  152.34 \\  
   21.73  & 456.31 &  24.03  &  266.20 &  17.55  &  159.32 \\   
   23.11  & 194.16 &   29.90   & 111.93 & 18.02 &   252.91\\   
   27.51  & 75.10 &   32.52   &  82.82 &  18.29  &  189.30  \\ 
   29.22  & 135.31 &  34.47  &  128.14 &  19.84  &  185.78 \\   
   36.44  & 126.24 &          &        &   21.47  &  144.24\\  
          &        &         &         &     21.50  &  162.64 \\ 
          &        &         &         &    22.86  &  289.28  \\
          &        &         &         &    24.18  &  141.95 \\ 
          &        &         &         &     24.30  &  143.60  \\
          &        &         &         &      24.59 &  324.46 \\  
          &         &       &          &      27.84 & 496.63  \\  
          &         &       &          &      29.42 & 158.36  \\  
          &         &       &          &      30.95 & 164.26 \\  
        \hline                                             
      \end{tabular} 
    \end{table*}

Infrared spectroscopy  is utilized for a characterization of  bulk cementitious systems, but results depend on  the variable chemical and structural composition
of these materials \citep{Yu1999,Garbev2007}.
Belite (Ca$_2$SiO$_4$) is one of the most important components of the cement paste. In  $\beta$-Ca$_2$SiO$_4$
a symmetrical stretching vibration of SiO$_4$ tetrahedra is assigned to bands measured at 11.57   and 11.74 $\mu$m \citep{Bensted1976}. 
In alite (Ca$_3$SiO$_5$), which  is also a major component of cement, weak bands exist at 6.42, 6.79 and 7.04 $\mu$m, and strong ones at 
10.64, 11.01, 11.32, 12.30, 18.98, and 22.08 $\mu$m \citep{Delgado1996}.
In the hydrated cement paste, where the ratio CaO/SiO$_2$  is 1.5, stronger bands appear at 30.03  (Ca-O lattice vibrations), 22.47 (O-Si-O bending vibrations in SiO$_4$), 14.88 (assigned to the bending of two connected SiO$_4$  tetrahedra), and 11.25 and 9.78 $\mu$m (both assigned to the symmetric stretching of SiO$_4$  tetrahedra) \citep{Garbev2007}.
Bands in the near-infrared region (at 6.1, 2.89, 2.81 and 2.76  $\mu$m) that correspond to the OH stretching vibrations in water and Ca(OH)$_2$ also exist in the spectra of hydrated cement paste. 
The most intense peak of cement is at 14.88 $\mu$m. This is the dominant and specific band  for all hydrated cement paste compositions, i.e., for the ratios CaO/SiO$_2$ between 0.2 and 1.5 \citep{Garbev2007}.

Spectral features at 9.2  and 21.7  $\mu$m  have been measured for calcium silicate smoke grains prepared in the laboratory by vapour phase condensation in the study of CaO and Ca(OH)$_2$ cosmic dust \citep{Kimura2005}.  Close bands exist for C2 (9.12 and 21.64 $\mu$m ) and C3 (8.99 and 21.50 $\mu$m).
A specific 14  $\mu$m  band of the bulk cement paste is present in the spectra of clusters.
We have found that bands at 14.44 and 14.55 $\mu$m (Fig. \ref{fig2}) exist for the most stable nanoparticle C1. This  should be compared with the strong band for C2  at 14.10 $\mu$m  (Fig. \ref{fig3}), and at 14.58 $\mu$m  for C3 (Fig. \ref{fig4}).
There is a strong band for C3 at 11.0{\bf 5}  $\mu$m, whereas in this region the weak band   at 11.46 $\mu$m exists 
for C2, and at  11.62  $\mu$m for C1.
Bands at 18 $\mu$m are also typical for amorphous cosmic silicates. For amorphous clusters bands appear at 18.38 $\mu$m (C2) and
at 18.02, 18.29, 18.58, and 18.99 $\mu$m (C3).
Bands at 18 $\mu$m  do not exist for the crystalline precursor C1.
While bands around (10--11) and 18 $\mu$m  appear in spectra of other silicates, features at 14 $\mu$m can be used to characterize cement nanoparticles.

Bands at 14 $\mu$m have been observed by the {\it Infrared Space Observatory} in the spectra
of objects from a sample of 17 oxygen-rich circumstellar dust shells  \citep{Molster2002a,Molster2002b}.
Weak features at 13.8 and 14.2 $\mu$m have been measured and classified as remaining bands or `(partly) instrumental'.
For example, bands at 13.5, 13.8, and 14.3 $\mu$m
in the spectrum of HD 45677 have been clearly detected.  Other examples of band observations  around 14 $\mu$m in the same sample are from the objects
HD 44179, Roberts 22, and IRC+10420.
However,  silicate bands at 10 and 11 $\mu$m have not been observed for  Roberts 22 \citep{Molster2002a}.
A silicate band at 18 $\mu$m exists for this object.
In contrast, silicate bands in both regions have been found in the spectra of HD 45677, HD 44179 and IRC+10420.
We expect that exact positions of cement bands could vary with the composition and size of nanoparticles.
To facilitate a comparison, in Table~\ref{tab2} 
the typical silicate and cement bands 
for the nanoparticles we studied are presented, together with the corresponding silicate and unidentified bands in the spectra of HD 45677, HD 44179 and IRC+10420 \citep{Molster2002a}. 
In Table~\ref{tab3}, bands above 19 $\mu$m are compared. Bands for cement nanoparticles of various compositions and sizes  cover all region of
observed features in these stellar objects.
In addition, HD 44179 was studied by {\it Spitzer} and unknown 
mid-infrared resonances in the 13--20 $\mu$m range were observed \citep{Kemper2005}. A broad peak at 13-17 $\mu$m was found and a mixture of Mg-Fe-oxides with unknown material was proposed as a carrier \citep{Kemper2005}. It is possible that cement nanoparticles contribute to this peak.

\begin{table*}
\centering
\caption{Silicate infrared bands at (9.7--11) and 18 $\mu$m, as well as 
specific cement bands around 14 $\mu$m, for nanoparticles C1 (crystalline precursor), C2 (amorphous), and C3 (amorphous).  Silicate 
and unidentified bands in the same spectral intervals
of three stellar objects \citep{Molster2002a} are also shown. Uncertain detections in the spectra of cosmic objects are marked by ``?''. 
}
\label{tab2}
  \begin{tabular}{ l| l |}
    \hline
    Object & Bands  ($\mu$m) \\
    \hline
    C1  &  10.26, 10.29, 10.40, 10.47, 10.63, 11.02, 11.11, 11.34, 11.59, 11.61, 11.62, 11.75, 14.44, 14.55 \\
    \hline
    C2  &  9.91, 10.18, 11.46, 11.73, 13.71, 14.10, 14.32, 18.38  \\
    \hline
    C3  &  9.74, 9.81, 10.27, 10.40, 11.05, 11.28, 11.60, 11.61, 13.63, 13.79, 14.58, 14.94, 18.02, 18.29, 18.58, 18.99 \\
    \hline
    HD 45677  &  9.81, 9.99, 10.57, 11.06, 11.50, 13.54, 13.77, 14.3, 18.06, 18.88\\
    \hline
    HD 44179  &  10.8?, 13.58, 14.19, 18.03, 18.97                       \\
    \hline
    IRC+10420 &  10.6, 10.7?, 11.04, 13.51, 13.76, 14.15, 18.2\\  
    \hline
  \end{tabular}
\end{table*}

\begin{table*}
\centering
\caption{Bands observed for stellar objects above 19  $\mu$m \citep{Molster2002a} and corresponding bands of cement nanoparticles.
Only bands with intensities above 10 km/mol are shown for C1, above 15 km/mol for C2, and above 50 km/mol for C3.
We expect that dust close to some oxygen rich stars consists of other silicates and oxides, as well as of the mixture of cement nanoparticles of various sizes, compositions, and infrared band positions.
}
\label{tab3}
  \begin{tabular}{ l| l |}
    \hline
    Object & Bands  ($\mu$m) \\
    \hline
    HD 45677  & 19.43, 20.67, 21.59, 22.41, 22.95, 23.67, 24.52, 25.04, 26.07, 26.93, 27.75, 28.30, 29.37,  \\
              & 30.61, 32.33, 32.79, 33.65, 34.5, 35.6, 38.21, 39.87, 40.64, 41.70, 42.95, 43.60, 44.90 \\
    \hline
    HD 44179  & 19.65, 20.77, 21.59, 22.86, 23.81, 24.65, 24.99, 26.08, 27.69, 28.26, 29.44, 30.77,  32.51,\\
              &   33.03, 33.70, 34.35, 35.28, 36.02, 36.53, 40.46, 43.5, 48.0, 56.6, 65.6, 69.02 \\
    \hline
    IRC+10420 & 20.70, 21.45, 22.96, 23.73, 24.16, 26.01, 27.48, 27.97, 32.56, 32.99, 33.54, 34.82, 35.95, \\ 
              & 40.37, 41.5, 43.00, 44.39, 47.83, 49.07, 54.9, 61.6 \\
     
    \hline
    C1        & 19.08, 21.22, 21.73, 21.96, 22.96, 23.11, 24.76, 26.25, 27.51, 28.16, 29.22, 31.22, 33.01, 33.60,  \\
              & 35.92, 36.44, 36.86, 38.76, 39.40, 46.31, 47.68, 48.39, 50.48, 50.89, 58.74, 60.51, 69.06  \\
    \hline
    C2        & 20.60, 20.92, 21.53, 21.64, 22.53, 22.90, 23.69, 24.03, 24.82, 25.49, 26.59, 29.42, 29.90,   \\
              & 32.16, 32.52, 33.05, 34.47, 37.08, 37.93, 39.55, 41.24, 43.43, 52.32, 57.04, 58.58 \\
    \hline
    C3        &  19.84, 20.68, 21.47, 21.50, 22.40, 22.86, 23.76, 24.18, 24.30, 24.59, 24.88, 25.96, 26.43,    \\   
              &  26.69,  27.68, 27.84, 28.98, 29.42, 30.95, 31.34, 32.14, 32.65, 33.35, 39.69, 41.09, 64.93 \\
    \hline
  \end{tabular}
\end{table*}

Oxygen is one of chemical elements for which depletion in the gas phase has been observed over many sight lines
\citep{Jenkins2009,Whittet2010,Jenkins2014}. It is known that in cosmic dust  O atoms exist in
silicates and metallic oxides. However, depletion of O atoms is much higher than,
for example, of Mg, Si, or Fe, and can be even greater 
by a factor of 16 \citep{Jenkins2009}. Existing dust grain composition models cannot explain this. 
The ratio of numbers of O atoms and Mg$+$Si atoms  even in the oxygen-rich MgSiO$_3$ is 1.5.
The ratio O/(Ca$+$Si) in our clusters
C1,  C2, and  C3 is 1.75, 1.44, and 3.25 respectively. Therefore, the number of oxygen atoms in cementitious nanoparticles is large, and could help in the  explanation of its depletion in the interstellar medium. A similar conclusion for the nanopyroxene cluster Mg$_4$Si$_4$O$_{12}$ has been reached by Goumans and Bromley
\citep{Goumans2011}.

\section{Conclusions}
\label{concl}
Silicates are one of main components of cosmic dust grains. However, their precise  chemical composition  is not known. To the pool of possible silicate materials in cosmic dust we have added cement. 
We have studied one crystalline and two amorphous 
Ca-Si-H-O clusters which represent the chemical composition and bonding of cement at the nanoscale.
Their infrared spectra are calculated, and it is found that bands are distributed over the entire infrared spectrum. 
We propose that the features at 14 $\mu$m, measured by the {\it Infrared Space
Observatory} in the dust shells of several oxygen-rich stars 
and currently classified as unidentified  \citep{Molster2002a,Molster2002b}, are the cement bands.
From the abundance of Ca and other chemical elements 
we do not expect that cement nanoparticles are a dominant species in cosmic dust.
However, cementitious nanoparticles, because of the many oxygen atoms they consist of, could act as an additional reservoir of O in cosmic dust and 
help in the solution of its depletion in the interstellar medium.
With a recent detection of Ca-rich supernovae that produce much larger amounts of calcium then previously expected \citep{Perets2010,Kawabata2010}, as well as after discoveries of water everywhere in Space by
the {\it Spitzer} and {\it Herschel}  missions \citep{vanDishoeck2013},  all the necessary ingredients for the formation of cement nanoparticles in space
are readily available.

\section*{Acknowledgments}
This work has been done using the computer cluster ``Isabella''  at the University of Zagreb Computing Centre SRCE.
GB acknowledges the support of the University of Zagreb research fund, grant ``Physics of Stars and Cosmic Dust''.
We are grateful to Goran Baranovi\' c for useful discussions.
We thank the editor,  Keith Smith, and the anonymous referee for their constructive comments and suggestions which have much
improved the clarity of this manuscript.

\bibliographystyle{mn2e} 
\bibliography{irdust}

\label{lastpage}
\end{document}